\documentclass[aps,prb,twocolumn]{revtex4-2}	

\usepackage{float}
\usepackage{hyperref}
\usepackage{graphicx,subfigure}
\usepackage{overpic}
\usepackage{cancel}
\usepackage{amsmath, amsfonts,amssymb}
\usepackage{bm}
\usepackage{sidecap}
\usepackage{color}
\usepackage{outlines}
\usepackage{soul}
\usepackage{dsfont}

\newcommand{\be}{\begin{equation}}
\newcommand{\ee}{\end{equation}}

\newcommand{\Tr}{{\rm Tr}}

\newcommand{\Ket}[1]{|#1\rangle}

\newcommand{\Avg}[1]{\langle #1 \rangle}

\newcommand{\norm}[1]{\lVert#1\rVert}

\newcommand{\cS}{\mathcal{S}}
\newcommand{\cB}{\mathcal{B}}
\newcommand{\cL}{\mathcal{L}}

\newcommand{\cV}{\mathcal{V}}
\newcommand{\cO}{\mathcal{O}}
\newcommand{\cH}{\mathcal{H}}

\begin{document}


\title{Quantifying the accuracy of steady states obtained from the Universal Lindblad Equation}
\author{
        Frederik Nathan$^1$ and 
        Mark S. Rudner$^2$
        }
\affiliation{
        $^1$Institute for Quantum Information and Matter, Caltech, Pasadena, California 91125, USA
\\
        $^2$Department of Physics, University of Washington, Seattle, Washington 98195, USA
}
\date{\today}
%

\begin{abstract}
We show that steady-state expectation values predicted by the universal Lindblad equation (ULE) are accurate up to  bounded corrections that scale linearly with the {effective} system-bath coupling, $\Gamma$ {(second order in the microscopic coupling)}. 
We also identify a near-identity, quasilocal ``memory-dressing'' transformation, {used during the derivation of the ULE}, whose {inverse can be applied to achieve} 
{\it relative} deviations {of observables} that generically scale to zero with $\Gamma$, even for  nonequilibrium currents whose  steady-state values themselves scale {to zero} with $\Gamma$. 
This result provides a solution to recently identified limitations on the accuracy of  Lindblad equations, which highlighted a potential for significant  relative errors in currents of conserved quantities. 
The transformation we identify allows for  high-fidelity computation of currents in the weak-coupling regime, ensuring thermodynamic consistency and  local conservation laws, while  retaining the stability and physicality of a Lindblad-form master equation.
\end{abstract}

\maketitle


The theoretical description {of open quantum systems is an important problem in many  fields of physics.
While early  groundwork  was  laid in the context of quantum optics and physical chemistry~\cite{Dirac_1927,Nakajima_1958,Zwanzig_1960,Redfield_1965,Davies_1974,Breuer}, recent advances in the control of complex  quantum systems necessitate the development of new theoretical techniques~\cite{Sieberer_2016,Kirsanskas_2018,Strathearn_2018,Nathan_2019,Kleinherbers_2020,McCauley2020,Mozgunov_2020,Nathan_2020b,Davidovic_2021,Somosa_2019,Becker_2021,Lerose_2021,Potts_2021,Mori_2022}. 

In this work we focus on the broadly relevant Markovian regime which emerges for weak system-bath coupling relative to  the  intrinsic energy scales of the baths. 
In this context, Lindblad equations play a key role by providing the most general form of Markovian  evolution that 
{is guaranteed to preserve the physicality of} solutions~\cite{Lindblad_1976, Gorini_1976}.
Formulating incoherent evolution in terms of randomly timed ``quantum jumps,'' Lindblad equations provide valuable physical insight and  admit stable and relatively efficient numerical solutions via stochastic evolution of wave functions (rather than density matrices)~\cite{Dalibard_1992,Carmichael_1993}. 
{The latter feature allows Lindblad equations} to be solved for systems of greater complexity than accessible with other approaches.

Historically, routes to {deriving} Lindblad equations {from the  evolution of a system coupled with its environment} have relied on a rotating wave approximation (RWA), which assumes the spectral gaps of the system to be larger than any decoherence rate~\cite{Davies_1974,Breuer}. 
While typically justified in 
{the traditional} contexts of quantum optics, the RWA is often not valid for complex quantum systems  emerging in 
{many} fields of interest such as hybrid quantum systems~\cite{kurizki_quantum_2015}, nonequilibrium many-body systems~\cite{kitamura_2019_floquet}, and platforms for large-scale quantum simulation and information processing~\cite{browaeys_many-body_2020,kjaergaard_2020_superconducting}.

Recently, several independent works discovered routes to circumvent the RWA~\cite{Kirsanskas_2018, Phd_thesis,Mozgunov_2020,Nathan_2020b,Davidovic_2021, Becker_2021, Potts_2021}, opening up {many of} the  
problems above for efficient study with  Lindblad equations~\cite{Kirsanskas_2018,Phd_thesis,Nathan_2019,Kleinherbers_2020,McCauley2020}. 
{Particular applications include, e.g., bath-induced localization in driven many-body systems~\cite{Maimburg_2021}, 
modelling of 
    qubit readout experiments~\cite{Mielke2021, aghaee2024interferometric} and 
{elucidating} the conditions of validity for the Landau-Lifshitz-Gilbert equation~\cite{garciagaitan2024fate}}.
In this work, we focus on the ``Universal Lindblad Equation'' (ULE), which was  derived in Ref.~\cite{Nathan_2020b} with rigorous bounds on approximation-induced errors; the bounds  are controlled by the product of a characteristic  {effective system-bath coupling} strength $\Gamma$ {(which is second-order in the microscopic system-bath coupling matrix elements)}, and a bath correlation time, $\tau$ [see Eq.~\eqref{eq:error_bound_def}].  


\begin{figure}[t]
        \includegraphics[width=0.99\columnwidth]{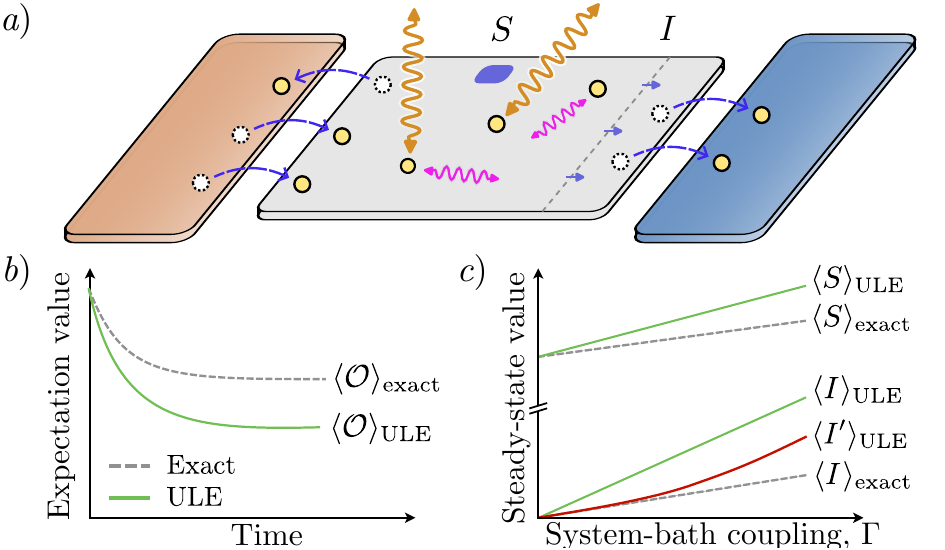}
        \caption{
        (a) We consider an open quantum system (gray) coupled to one or more baths, such as, e.g.,  phonons/photons (purple/orange), or particle reservoirs (red/blue). 
        We distinguish generic  observables, $S$, such as particle or spin densities, from currents of conserved quantities, $I$.
        (b) The expectation value of any observable $\mathcal O = S,I$ generically converges to a unique value at long times, $\langle  \mathcal O\rangle_{\rm exact}$, regardless of initialization.
        (c) We  show that the corresponding steady-state value resulting from the universal Lindblad equation (ULE), $\langle  \mathcal O\rangle_{\rm ULE}$, deviates from $\langle  \mathcal O\rangle_{\rm exact}$ by a value that scales to zero with the system-bath coupling, $\Gamma$.
        In the limit, $\Gamma \to 0$, the ULE hence accurately captures steady-state values that remain finite, as occurs for  generic observables  $S$.
        For  nonequilibrium currents, $I$, whose  steady-state values  vanish in the limit $\Gamma\to 0$, an accurate steady-state value  can be obtained from a ``dressed'' current $I'$ that results from  applying a near-identity, quasilocal   transformation to $I$ [Eqs.~(\ref{eq:M_timedomain}),~(\ref{eq:Obs_corrected})].
        }        \label{fig:figure_1}
        \vspace{-0.2 in}
\end{figure}

In parallel with the advances above, several groups recently   explored the nontrivial tradeoffs of studying
nonequilibrium quantum systems with Lindblad equations.
Even away from equilibrium, an open quantum system, e.g., as depicted in Fig.~\ref{fig:figure_1}(a), generically converges to a  unique {\it non-equilibrium steady state}, as schematically depicted for some given observable $\mathcal O$  by the dashed black curve
  in Fig.~\ref{fig:figure_1}(b).
The authors of Ref.~\cite{Tupkary_2021} showed that the steady state of any Lindblad equation
must deviate from the exact steady state by a correction linear in the {effective} system-bath coupling [the green curve in Fig.~\ref{fig:figure_1}(b) 
depicts the flow of $\Avg{\mathcal O}$ predicted by the ULE].
While the steady state of a Lindblad equation can approach the exact steady state in the weak-coupling limit, nonequilibrium currents such as, 
e.g., particle or heat
currents, scale linearly with system-bath coupling. 
{Consequently}, their {\it relative deviations} 
generically remain finite {as the coupling is taken to zero}.
This can lead to apparent violations of the second law of thermodynamics~\cite{Kirsanskas_2018,Potts_2021} and local conservation laws~\cite{Tupkary_2021}.

To enable further progress in the field, here we quantify the fidelity of steady states resulting from the ULE, and demonstrate how
the apparent inconsistencies above can be resolved.
In Sec.~\ref{sec:summary} we summarize our main results.
In Sec.~\ref{sec:setup} we review the problem addressed by our work: 
quantifying the accuracy of steady states of the ULE. 
In Sec.~\ref{sec:ule_review} we revisit the derivation and key features of  the ULE relevant for the problem.
In Sec.~\ref{sec:steady_state_fidelity} we present our solution via rigorous bounds on the deviations of {observables computed with} the ULE steady state 
{compared with those obtained from} the exact steady state. 
In Sec.~\ref{sec:demonstration} we demonstrate our results in numerical simulations of magnon transport in a spin chain. 
We conclude with a discussion in Sec.~\ref{sec:discussion}.
Details of derivations and supplemental data are provided in Appendices.

\section{Summary of main results}\label{sec:summary}
We report two main results,  summarized in Fig.~\ref{fig:figure_1}(c): 
\begin{enumerate}
\item First, we show that in the limit $\Gamma \tau \to 0$ the  ULE  faithfully  captures the  steady-state values  of generic   observables in {\it absolute} scale: for a generic observable,  the   steady-state value predicted by the ULE deviates from the true steady-state value by an amount  that  scales to zero with $\Gamma\tau$  [Eq.~\eqref{eq:steady_state_without_lft}]. 

\item Second, we demonstrate that  steady-state expectation values with vanishing {\it relative} deviations in the limit $\Gamma \tau \to 0$ can be obtained   
{by applying the inverse of} a quasilocal, near-identity     ``{memory-dressing}'' transformation that was employed to obtain the ULE~\cite{Nathan_2020b}  [see Eqs.~\eqref{eq:M_timedomain} and \eqref{eq:Obs_corrected} below]. 
The transformation   only affects observables with support near regions connected to  baths.
This result enables the computation of accurate and thermodynamically-consistent nonequilibrium currents which obey   local conservation laws.
Our analysis is supported by numerical simulations (Fig.~\ref{fig:fig2}).
\end{enumerate}

{Before embarking on the technical discussion, we offer a few clarifying remarks about the nature of 
our results.
As a Lindblad equation, the ULE provides a stable, positivity preserving map for the evolution of an open system's density matrix.
The ULE can therefore be used to study the trajectories and steady states of open quantum systems, by going to the long-time limit.
After the positivity-preserving evolution or corresponding steady state is computed from the ULE, the inverse of the memory dressing transformation can be applied to remove some of the deviation between the steady-state solution of the ULE and the  exact steady state  of the system.
The previously identified trade-offs or apparent limitations of Lindblad equations~\cite{Tupkary_2021} are resolved by the fact that the inverse of the memory dressing transformation breaks the positivity of the density matrix {\it in a small and controlled way, bounded by $\Gamma\tau$}, which brings it and associated observables closer to their true values. 
Since the transformation is applied {\it after} the evolution, the stability of the Lindblad evolution and applicability of quantum trajectory approaches are retained.
}

{We also emphasize that 
the role of the memory dressing transformation can often be neglected:} it is only required for computing nonequilibrium currents or other quantities whose steady-state values vanish in absolute terms in the weak-coupling limit (i.e., {\it not} for quantities whose expectation values remain finite).
{Moreover, as we demonstrate  below,} even for observables with vanishing steady-state values in the weak-coupling limit, the memory dressing transformation is only needed 
{if their support lies} 
close to points of contact with baths.
}


\section{Formulation of problem}
\label{sec:setup}
We consider a  quantum system  with Hamiltonian $H_{\mathcal S}$, where one or more degrees of freedom $\{X_\alpha\}$ are  coupled to degrees of freedom  in the surrounding environment, or ``bath,'' $\{B_\alpha\}$, such as, e.g.,  photonic or phononic  modes, or creation/annihilation operators  of particles in attached  reservoirs {[see Fig.~\ref{fig:figure_1}(a)]}. 
The environment can, e.g.,  consist of  several baths out of mutual equilibrium, where different subsets of  $\{B_\alpha\}$ act on distinct baths. 
The combined system has Hamiltonian 
\begin{equation}
        H_{\cS\cB} = H_{\mathcal S} + H_{\mathcal B} + \sqrt{\gamma}\sum_{\alpha=1}^N X_\alpha B_\alpha,
\end{equation}
where  $H_{\cB}$ is the Hamiltonian of the bath, and we introduced
 $\gamma$ as a redundant energy scale  parameterizing the system-bath coupling strength.
We take each $X_\alpha$ to have unit spectral norm and take the environment to  be in a state where each $B_\alpha$ has expectation value zero~\cite{sbnorm}.

We consider the broadly relevant case where  the environment is Gaussian (e.g., comprised  of free bosonic or fermionic modes). 
A Gaussian environment is  fully characterized by its two-point correlation function $J_{\alpha\beta}(t) = \langle B_\alpha(t)B_\beta(0)\rangle$ with $B_\alpha(t)\equiv e^{iH_{\mathcal B}t} B_\alpha e^{-iH_{\mathcal B}t}$. 
Our discussion considers two additional   objects  which equivalently  encode the properties of the environment:  the  bath spectral function, $J_{\alpha\beta}(\omega)\equiv \int \frac{dt}{2\pi}  e^{-i\omega t}J_{\alpha\beta}(t)$, and the ``jump correlator'' $g_{\alpha\beta}(t) = \int d\omega\, e^{-i\omega t}[\sqrt{J(\omega)/2\pi}]_{\alpha\beta}$, with $\sqrt{\cdot}$ denoting the matrix square root. 
 
Observables of the system are determined  by the reduced density matrix
\begin{equation}
\label{eq:rho_def} \rho(t)= \Tr_{\mathcal B}[\rho_{\mathcal S\mathcal B}(t)].
\end{equation}
Here $\rho_{\mathcal S\mathcal B}(t)$ denotes the exact density matrix of the combined system, whose evolution is described by $\partial_t \rho_{\cS\cB}(t) = -i[H_{\cS\cB},\rho_{\cS\cB}(t)]$.
For thermodynamically large environments, $\rho(t)$ generically goes to a unique steady state in the long-time limit regardless of the initial state of the system, $\rho_0$~\cite{unique_fixed_point}:
\be 
\bar \rho_{\rm exact} \equiv \lim_{t \to \infty} \rho(t).
\ee 
Away from   equilibrium, $\bar \rho_{\rm exact}$ depends on the details of the environment and the system-bath coupling. 
Such nonequilibrium steady states can be found by establishing approximate equations of motion for $\rho(t)$ and seeking their steady state solutions.

In the 
{regime} where the {effective} system-bath coupling is  small relative to the inverse  correlation time  of the bath, it was recently 
{shown that the system's evolution is well-described by the ULE~\cite{Phd_thesis,Nathan_2020b,Davidovic_2021}}: 
\be
\partial_t \rho
\approx 
i[\rho, H_{\mathcal{S}}+ H_{\rm LS}] 
+ \sum_{\alpha=1}^{N}\left( L_\alpha \rho  L_\alpha^\dagger 
- \tfrac{1}{2}\{ L_\alpha^\dagger  L_\alpha ,\rho\} \right),
\label{eq:ule_liouvillian}
\ee
where $L_{\alpha} = \sum_{mn\beta}[\sqrt{2\pi \gamma J(E_m-E_n)}]_{\alpha\beta} |m\rangle\langle m|X_\beta|n\rangle \langle n|$ with $H_{\mathcal S}|n\rangle = E_n|n\rangle$.
The Hermitian Lamb shift  $H_{\rm LS}$ 
effectively renormalizes $H_{\mathcal S}$; its form is given in Ref.~\cite{Nathan_2020b}. 
Our goal  is to characterize the fidelity of the steady state of the ULE, 
\be 
\bar \rho _{\rm ULE} \equiv \lim_{t\to\infty} e^{\mathcal L_{\rm ULE}t}[\rho_0].
\ee 
Here  $\cL_{\rm ULE}$ denotes the ULE Liouvillian, whose action on an arbitrary density matrix $\rho$ is given by the right-hand side of Eq.~\eqref{eq:ule_liouvillian}.
Note that  $\cL_{\rm ULE} [\bar \rho_{\rm ULE}] = 0$.


\section{ Review of the ULE}
\label{sec:ule_review}
We first 
review the key features
of the  ULE~\cite{Nathan_2020b}. 

\subsection{Derivation of the ULE}
The validity  of the ULE is controlled by two timescales, $\Gamma^{-1}$ and $\tau$, given by 
\be 
    \Gamma = 4\gamma  \left[\int_{-\infty}^\infty dt\, \norm{g(t)}_{2,1}\right]^2,\quad \tau = \frac{ \int_{-\infty}^\infty dt\, \norm{g(t) t}_{2,1}}{\int_{-\infty}^\infty dt\, \norm{g(t)}_{2,1}},
    \label{eq:error_bound_def}
\ee
where $\norm{g}_{2,1}\equiv\sum_{\beta}(\sum_\alpha |g_{\alpha\beta}|^2)^{1/2}$. 
The rate 
$\Gamma$ bounds the exact rate of evolution induced by the bath: 
$
    \norm{\partial_t \rho~-~i[\rho,H_S]} 
    \leq \Gamma/2
$,
where $\norm{\cdot}$ denotes the trace norm~\cite{trace_norm_bound,Nathan_2020b}. 
The timescale $\tau$ is a characteristic correlation time of the bath; the dimensionless number $\Gamma \tau$ controls the approximations involved in deriving the ULE (as well as the Bloch-Redfield equation~\cite{Mozgunov_2020,Nathan_2020b}).

To derive the ULE, {starting from Eq.~(\ref{eq:rho_def})} we first expand $\partial_t\rho$ to leading order in $\gamma$, yielding 
    $\partial_t{\rho} (t)= {\mathcal L}_{\rm BR}[\rho(t)] +   {\xi}_{\rm BR}(t)$~\cite{Redfield_1965,Mozgunov_2020,Nathan_2020b},
where ${\mathcal L}_{\rm BR}$ denotes the Bloch-Redfield (BR) Liouvillian~\cite{br_liouvillian}.
The residual  ${\xi}_{\rm BR}(t)$ captures the difference between the exact value of $\partial_t{\rho}$ and that resulting from the BR 
equation~\cite{Nathan_2020b}. 
{Refs.~\cite{Nathan_2020b,Mozgunov_2020} recently established that} $\norm{{\xi}_{\rm BR}(t)}\leq 2\Gamma^2 \tau$~\cite{trace_norm_bound}.

We obtain the ULE by transforming to a weakly ``memory-dressed'' frame  of operator space 
using a near-identity quasilocal superoperator, $\cV$:
\be 
\rho'=    \mathcal V[{\rho}].
\label{eq:lindblad_transformation}
\ee 
We provide an explicit expression for {$\cV$} in Eq.~\eqref{eq:M_timedomain} below. 
By design, $\cV$ preserves the Hermiticity and trace of $\rho$, and satisfies 
$-i[\mathcal V,\cH_{\cS}] ={\mathcal L}_{{\rm ULE}}-  {\mathcal L}_{{\rm BR}}$~\cite{Nathan_2020b}, where $\cH_{\cS}[\rho]\equiv [H_S,\rho]$.
Evaluating $\partial_t \rho '(t)$  gives 
\be 
\partial_t \rho ' (t)= \mathcal L_{\rm ULE}[\rho'(t)] 
+{\xi}(t),
\label{eq:ule_exact}\ee
{where the residual 
$\xi(t)$  satisfies  
$\norm{\xi(t)}\leq 2\Gamma^2 \tau$~\cite{ule_residual_proof}. 
The ULE is obtained by neglecting  $\xi(t)$, which is subleading in $\Gamma \tau$, since ${\mathcal{L}}_{\rm ULE}$ consists of terms of order $\Gamma$ and $\Gamma^0$}.

Ref.~\cite{Nathan_2020b} showed that $\norm{  \mathcal V-{\bf 1}}_{\rm SO}\leq \Gamma \tau$,  where $\norm{\mathcal A}_{\rm SO} = \sup_{{\mathcal O}} \norm{\mathcal A[{\mathcal O}]}/\norm{{\mathcal O}}$.
As a result, $\rho$ and $\rho'$ nearly coincide when $\Gamma \tau \ll 1$:
\be 
\norm{\rho'-\rho}\leq \Gamma \tau. 
\label{eq:rho_rho'_relation}
\ee
Note that  both $\norm{\rho-\rho'}\ll 1$ and  $\xi$ are negligible in the limit $\Gamma \tau \ll 1 $.
In this sense, the ULE is accurate when $\Gamma \tau \ll 1$.

\subsection{Memory-dressing transformation}
\label{sec:frame_transformation}
The  transformation to the memory-dressed frame, $\cV$, is given by $ {\bf 1} +\delta \cV$, with
\begin{align}
\delta {\mathcal V}[\rho ] &= 
\gamma\! \int_{-\infty}^\infty\!\!\! dt dt' [\tilde X_{\alpha}(t) , \rho\tilde X_{\beta}(t')] f_{\alpha\beta}(t,t')  + {\rm h.c}.,
\label{eq:M_timedomain}
\end{align}
{where $\tilde X_\alpha(t) = e^{iH_{\cS} t}X_\alpha e^{-iH_{\cS} t}$, the indices $\alpha$ and $\beta$ are implicitly summed over, and} 
$f_{\alpha\beta}(t,t') \equiv \sum_{\lambda}\int_{-\infty}^\infty \!\! ds\, g_{\alpha\lambda}^*(t-s)g_{\lambda\beta}^*(s-t')\theta(t-t')[\theta(t)-\theta(s)]$. 
{In App.~\ref{app:lindblad_frame_transf_expr} we provide alternative expressions for}  $\mathcal V$ in terms of the eigenstates  of $H_{\mathcal S}$, and via an iterative  expansion.
Whereas the standard Markov approximation made en route to the Bloch-Redfield equation~\cite{Breuer} amounts to approximating the density matrix as constant within the brief window of nonzero bath-correlations, 
passing to the memory-dressed frame effectively incorporates a more intricate averaging of the density matrix over the window of decaying correlations in the bath.

The memory frame transformation $\cV$ can be shown to  induce $O(\Gamma{\tau})$
corrections  to $\rho$, and  can thus typically be ignored in the limit  $\Gamma \tau \ll 1$~\cite{Nathan_2020b}.
However, the corrections  {\it are} relevant for observables whose steady-state values scale with $\Gamma$, such as nonequilibrium currents.
We demonstrate below how  {to obtain} negligible relative errors of these quantities by applying the inverse of the memory dressing transformation to the ULE steady state,  i.e., by computing their expectation values with the operator
$\cV^{-1}[\bar \rho_{\rm ULE}]$.
Using $\norm{\delta \cV}_{\rm SO}\leq\Gamma\tau$, Taylor expansion yields
$\cV^{-1} = 1-\delta\cV + \delta ^{(2)} \cV $, where, for $\Gamma \tau< 1$, $\norm{\delta^{(2)}\cV }_{\rm SO}\leq (\Gamma\tau)^2/(1-\Gamma \tau)$. 

Importantly,  $\cV$ only affects observables with support near regions connected to baths:
firstly,  the integrand defining $f_{\alpha\beta}$ below Eq.~\eqref{eq:M_timedomain} is nonzero only for $t > 0, s < 0$ and vice versa.
For a bath with a smooth spectral function (i.e., finite memory), $g(t)$  decays to zero for large $t$~\cite{Nathan_2020b} with a characteristic decay time $\tau_g$.
Hence, $f(t,t')$ also decays with a characteristic timescale of order $\tau_g$ for large $|t|$ or $|t'|$. 
Moreover, $\tilde X_\alpha(t)$ only has support  within a distance $v_{\rm LR} t$ from the support of $X_\alpha$, where $v_{\rm LR}$ denotes the system's Lieb-Robinson velocity.
Thus, $\cV$ only significantly affects    operators with support within a  distance  $\ell_{\mathcal{V}} \sim  v_{\rm LR}\tau _g$ from 
regions connected to baths. 

\newcommand{\lule}{\mathcal{L}_{\rm ULE}}

\section{fidelity of ULE steady state observables}
 \label{sec:steady_state_fidelity}
 Here we  bound  the differences between  the exact steady-state value of a given observable, and  the steady-state values  
 {resulting from} the ULE with and without accounting for the memory-dressing transformation.
 In the following, we will use $\langle \mathcal O\rangle_{\rm exact}$ and $\langle \mathcal O\rangle_{\rm ULE}$ to denote expectation values of the observable $\mathcal O$ in the exact steady-state and the ULE steady state, respectively:
\be
\langle  \mathcal O\rangle_{\rm exact} \equiv   \Tr[\bar \rho _{\rm exact}\mathcal O],
\ \  
\langle  \mathcal O\rangle_{\rm ULE}   \equiv   \Tr[\bar \rho _{\rm ULE}\mathcal O].
\label{eq:Obs_uncorrected}
\ee

To facilitate the comparison, 
we first translate $\langle  \mathcal O\rangle_{\rm exact}$ 
to the memory-dressed frame via $\Avg{\cO}_{\rm exact} =\Tr[\cO'\bar \rho'_{\rm exact}]$, 
where  $\bar \rho'_{\rm exact}\equiv\cV[\bar\rho_{\rm exact}]$, and~\cite{superoperator_adjoint}
\be
 {\cO}' \equiv \cV^{-1\dagger}[\cO].
\label{eq:Obs_corrected}
\ee
Next, {by setting  $\partial_t\rho' = 0$ in Eq.~(\ref{eq:ule_exact})}, note that
\be
\bar \rho'_{\rm exact} = \bar \rho_{\rm ULE}+ \mathcal L_{\rm ULE}^{-1}[\bar \xi],
\ee
where $\bar\xi =\lim_{t\to \infty}\xi(t)$; 
see Appendix~\ref{app:exact_and_ule} for derivation.
Combining these  results  with $\norm{\bar \xi}\leq  2\Gamma^2\tau$, we conclude
\be 
\left|\langle  {\cO}'\rangle_{\rm ULE} - \langle  \cO \rangle_{\rm exact}  \right| \leq C_\cO\Gamma\tau  |\langle  \cO'\rangle_{\rm ULE} |,
\label{eq:ltf_expval}
\ee 
where 
$
C_\cO \equiv 2 \Gamma\norm{\cL_{\rm ULE}^{\dagger -1}[\cO'-\langle \cO'\rangle _{\rm ULE}]}/|\langle \cO'\rangle_{\rm ULE}|$.
Crucially, as we show in Appendix~\ref{app:co_scaling}, $C_\cO$ is finite   for generic $\cO$, and remains so 
as $\Gamma \to 0$, including for currents whose steady-state values scale to zero with $\Gamma$~\cite{infca}.
Thus, the relative deviation of   {$\langle \mathcal O'\rangle _{\rm ULE} = {\rm Tr}\{\mathcal{V}^{-1}[\bar{\rho}_{\rm ULE}]\mathcal{O}\}$} generically vanishes in the limit $\Gamma \tau \ll 1$. 

The  transformation to the memory-dressed frame  can typically be neglected if $\cO$ has a finite steady-state value in the limit $\Gamma\tau \ll 1$. 
Specifically, using Eq.~\eqref{eq:ltf_expval}  along with $|\langle  \cO'\rangle_{\rm ULE} -\langle   \cO\rangle_{\rm ULE}| \leq \frac{\Gamma \tau}{1-\Gamma \tau} \norm{\cO}_2$ (where $\norm{\cdot}_2$ denotes the maximal singular value norm)~\cite{sing_val_result}  gives
\be 
|\langle  \cO\rangle_{\rm ULE}
- \langle {\cO}\rangle_{\rm exact}| 
\leq   
C_\cO\Gamma \tau | \langle \cO\rangle_{\rm ULE}|
+  \frac{\Gamma \tau+ C_\cO  \Gamma^2\tau^2}{1-\Gamma \tau}\norm{\cO}_2.  
\label{eq:steady_state_without_lft}
\ee
The right hand side of Eq.~(\ref{eq:steady_state_without_lft}) scales to zero when $\Gamma \tau\to 0$, implying that the {absolute} deviation of $\langle \cO\rangle_{\rm ULE}$ vanishes in this limit. 
The magnitude of the correction from the memory-dressing transformation (second term  above) moreover decays to zero with the distance from $\cO$ to the bath, {over the characteristic scale $\ell_{\mathcal V}$}, due to the quasilocality of $\mathcal V$. 
Thus, in the weak-coupling limit, the correction from $\cV$ is only relevant if $\lim_{\gamma\to 0}\cO= 0$ {\it and} $\cO$ has  support within a distance $\ell_{\mathcal{V}}$ from regions connected to baths. 

Equations \eqref{eq:ltf_expval} and \eqref{eq:steady_state_without_lft} 
 are our main results: Eq.~\eqref{eq:steady_state_without_lft} 
 demonstrates that steady-state expectation values of observables computed from the ULE have vanishing {\it absolute} deviations from their exact counterparts in the weak-coupling limit. 
 Equation~\eqref{eq:ltf_expval} 
 shows that ULE steady-state observables {are guaranteed to} have vanishing {\it relative} deviations when 
 {the transformation to the memory-dressed frame 
 is accounted for by using the memory-dressed observable, $\langle \mathcal{O}'\rangle_{\rm ULE} = {\rm Tr}[\bar{\rho}_{\rm ULE} \mathcal{O}']$,  rather than the bare observable,  $\langle \mathcal{O}\rangle_{\rm ULE} = {\rm Tr}[\bar{\rho}_{\rm ULE} \mathcal{O}]$. 
We emphasize that the bounds in Eqs.~\eqref{eq:ltf_expval} and \eqref{eq:steady_state_without_lft} are loose, and define   
``worst case'' deviations. 
We expect that actual deviations can often 
be much smaller.

\section{Numerical demonstration}
\label{sec:demonstration}
We  demonstrate our results in simulations of magnon transport in a spin chain. 
We consider a chain of $9$ sites with Hamiltonian $H_\mathcal{S} = \sum_{n=1}^{8} g(\sigma_{n}^x\sigma_{n+1}^x + \sigma_{n}^y\sigma_{n+1}^y+ \Delta \sigma_{n}^z\sigma_{n+1}^z) + \sum_{n=1}^9 h_n \sigma_n^z
$, where $\sigma_n^\alpha$ is the $\alpha = \{x,y,z\}$ Pauli operator on spin $n$; the same model was studied in Ref.~\cite{Tupkary_2021}.
We couple $\sigma_1^x$ and $\sigma_9^x$ to two Gaussian baths through the terms $
\sqrt{\gamma}\sigma_1^x B_1 + \sqrt{\gamma}\sigma_9^x B_2$.
The baths are Ohmic, with temperatures $T_1$ and $T_2$.
The  bath spectral function  reads $J_{\alpha\beta}(\omega) = \delta_{\alpha\beta}\frac{\omega e^{-\omega^2/\Lambda^2}}{\omega_0}n_{\rm B}(\omega/k_{\rm B}T_\alpha)$, with $\delta_{\alpha\beta}$ the Kronecker delta, $\omega_0$  a normalizing frequency, $\Lambda$  a high-frequency cutoff, $n_{\rm B}$  the Bose-Einstein distribution, and  $k_{\rm B}$  the Boltzmann constant. 
We simulate the model with $k_{\rm B}T_1 = g$, $k_{\rm B}T_2 = 6g$, $\Delta = 1.4g$, $\Lambda = 8g$, and $h_n = \frac{2}{3}(n-5)  g$, describing a uniform field gradient.

\begin{figure}
\includegraphics[width=0.99\columnwidth]{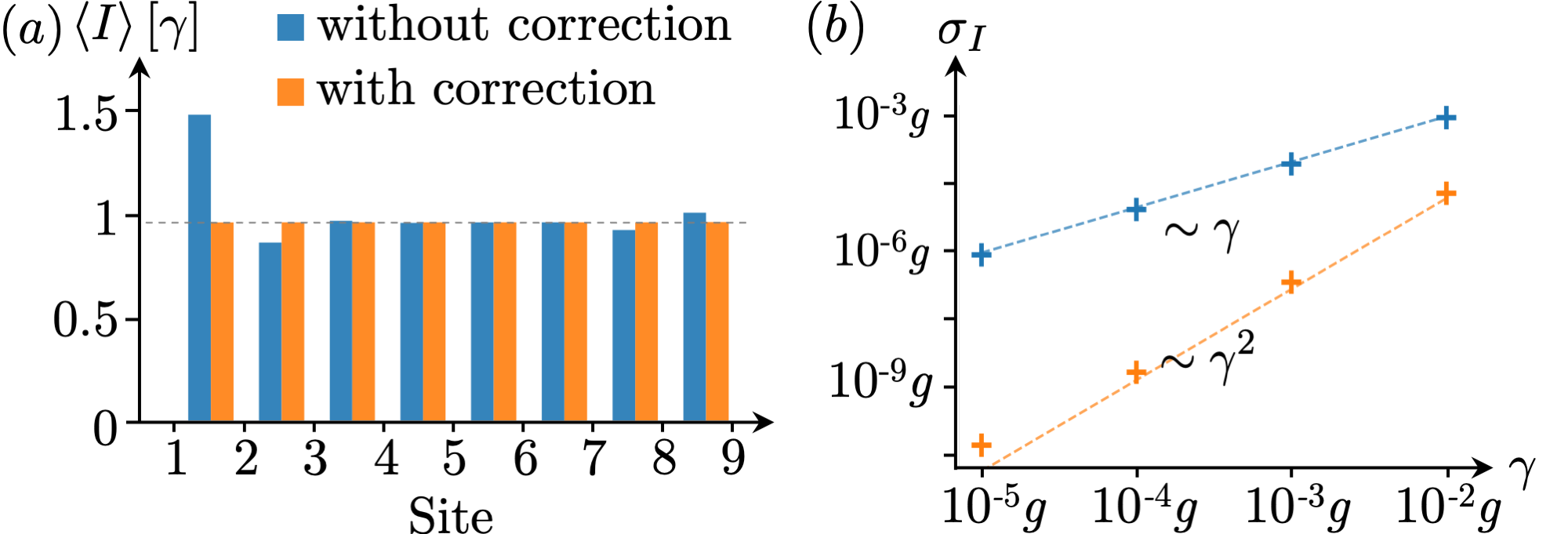}
\caption{
Bare and dressed steady-state bond currents in a $9$-site spin chain connected to two baths of unequal temperatures; see text for model details. 
(a) The bare bond currents vary significantly from site to site (blue bars), apparently violating current conservation.
The dressed currents (orange bars) are  near-uniform.
Note  the negligible effect of the memory dressing transformation deep in the bulk.
(b) Scaling of the standard deviation of the bond  currents  in the $9$ site chain, $\sigma_I$, as a function of 
$\gamma$ (crosses). 
}
\label{fig:fig2}
\vspace{-0.2 in}
\end{figure}

The system's spectrum, $\{E_{m}-E_{n}\}$, has mean adjacent-value spacing $ 4.9\cdot 10^{-5} g$, while the mean decay rate  prescribed by   Fermi's golden rule is given by $\frac{2\pi \gamma}{D}\sum_{n,m=1}^{D}\big[J_1(E_n-E_m)|\langle n|\sigma^x_1|m\rangle |^2+ J_2(E_n-E_m)|\langle n|\sigma^x_9|m\rangle |^2\big] \approx  23 \gamma$, where $D = 2^9$ is the system's Hilbert space dimension. 
The commonly used  Davies  equation~\cite{Davies_1974,GardinerZoller} (based on the RWA) is thus expected to be valid  when $\gamma \ll  2.2\cdot 10^{-6}g$.
On the other hand, explicit calculation shows $\Gamma \tau\approx  31 \gamma/g$, implying that  the approximations leading to the ULE are  justified when  $\gamma \ll 0.03 g$.

Microscopically, the magnon current on the bond from site $n$ to site $n+1$ is given by 
$I_{n+1,n}= (4i g \sigma_{n+1}^+ \sigma_{n}^- + {\rm h.c.})$.
Since the exact evolution of the combined system yields $\partial_t \sigma_n^z  = I_{{n,n-1}} -  I_{n+1,n}$ for  $1<n<9$,
$\langle I_{n,n+1}\rangle_{\rm exact}$ must be uniform throughout the system.
On the other hand, $\langle  I_{n+1,n}\rangle_{\rm ULE}$ may be non-uniform  near the baths, because   the quasilocal Lamb shift and jump operators of the ULE modify the equation of motion for $\sigma_n^z(t)$. 
Nontrivially, Eq.~(\ref{eq:ltf_expval}) predicts  that the {\it dressed} bond currents, $\langle  I'_{n,n+1}\rangle_{\rm ULE}$,  are homogeneous throughout the  system, up to   relative deviations that scale to zero with $\gamma$. 
This feature thus serves as a consistency check of our analysis and indicator of the fidelity of the bond currents  obtained with the ULE~\cite{fn_bond_current_uniformity}.

Fig.~\ref{fig:fig2}(a) shows the steady state values of  $\{\Avg{{I}'_{n+1,n}}_{\rm ULE}\}$ and $\{\Avg{{I}_{n+1,n}}_{\rm ULE}\}$~\cite{invlft},   obtained  for $\gamma = 10^{-3}g$. 
Near-identical data are found for   all probed values of $\gamma$ smaller than $10^{-2}g$.
Evidently, the bare currents exhibit significant non-uniformities near the ends of the chain, while the dressed currents are nearly uniform. 
These deviation are consistent with  expected subleading   $\mathcal O(\gamma^2)$ corrections:
in Fig.~\ref{fig:fig2}b, we  show the $\gamma$-scaling of the standard deviations of    the bare and dressed current. 
While the standard deviation of  the bare currents  scales   linearly  with $\gamma$, that of the dressed currents
 scales {\it quadratically} with $\gamma$.
Deep in the bulk, the bare and dressed bond currents in Fig.~\ref{fig:fig2}(a) are near-identical, indicating that the memory dressing transformation only affects observables in vicinities of baths.
In Appendix~\ref{app:magnetization} we  provide data confirming that the  memory dressing transformation is not needed to accurately capture the expectation values of 
 $\{\sigma^z_n\}$,  which remain finite when $\gamma \to 0$.

We finally  bound   the deviation of the magnon current precicted by the ULE via Eq.~\eqref{eq:ltf_expval}. We obtain $C_{I_{1,2}}\approx  1300$ for multiple values of   $\gamma $ between $10^{-5}g$ and $10^{-2}g$ (indicating that $\lim_{\gamma \to 0}C_{I_{1,2}}$  is finite), and   $\Gamma \tau \approx 31\gamma/g$. Hence the current predicted by the ULE is {\it guaranteed} to be accurate (have relative deviation  much smaller than $1$) when $\gamma \ll 2.5 \cdot 10^{-5}g$.
We expect $\langle I'_{\rm 1,2}\rangle_{\rm ULE}$ to be accurate   for values of $\gamma$ above  this  rigorous, but loose, bound: 
the standard deviation of the dressed bond currents is of order $1\%$  even for $\gamma \approx 10^{-2}g$  [see Fig.~\ref{fig:fig2}(b)], suggesting that the corrected ULE observables may remain accurate up to this value.


\section{Discussion}
\label{sec:discussion}
In this work, we   established upper bounds on the steady-state {deviations of observables computed from} the ULE. Our analysis revealed two results: (1) We demonstrated that   the ULE yields accurate steady-state values of observables in the weak-coupling limit, in {\it absolute} terms, and 
(2) we  showed that  accurate steady-state values in  {\it relative} terms can be achieved by applying a near-identity, quasilocal  memory-dressing transformation   to  the observables in question  [Eqs.~\eqref{eq:lindblad_transformation}-\eqref{eq:M_timedomain}].
The transformation is only needed for}  current-like observables  with support close to regions connected to  baths.

Even for   situations where   the  memory-dressing transformation is needed, the {advantageous characteristics of the ULE remain intact.} 
The transformation 
{is only} applied {\it after} evolving the system with the ULE until it reaches its steady state. 
When applied, the transformation only induces a small, bounded correction {to the state or observables}. 
As a result,  the procedure remains stable and solvable with stochastic evolution methods (and is hence parallelizable and efficient). 
The residual deviations are comparable to those expected for the Bloch-Redfield equation (BRE), while avoiding the potential instabilities of the BRE and enabling the use of efficient computational schemes as mentioned above (see, e.g., Ref.~\cite{breuer2024benchmarking}, where extensive benchmarking was recently performed for a dissipative harmonic oscillator, albeit without the memory-dressing transformation.)

Our results facilitate 
accurate computation of non-equilibrium currents in the weak-coupling limit via Lindblad equations, ensuring thermodynamic consistency and local conservation laws. Thereby, we demonstrated how to overcome  the fundamental limitations  intrinsic to {any} Lindblad equation that were pointed out in Refs.~\cite{Potts_2021, Tupkary_2021}.  

Important future directions 
will be to obtain tighter error bounds and to extend the proven regime of validity of the ULE, e.g., by formulating conditions in terms of the spectral range of the system~\cite{Davidovic_2021}.
Other interesting directions involve extending our bounds on steady-state deviations to driven quantum systems, and exploring routes to computing nonequilibrium currents which circumvent the memory dressing transformation.
More broadly, our results  enable  systematic   studies   of  mixed coherent/incoherent evolution  in  complex  quantum systems, relevant, e.g.,  for characterizing coherences in noisy devices  for quantum information processing.


{\it Acknowledgments ---}
We would like to thank A. Dhar, G. Kirsanskas, M. Kulkarni, M. Leijnse, A. Purkayastha, and D. Tupkary for helpful and clarifying discussions and for pointing out the limitations of the ULE (and Lindblad equations in general), which we address in the present work. F.N. and M.R.
gratefully acknowledge the support of Villum Foundation,
the European Research Council (ERC) under the European
Union Horizon 2020 Research and Innovation Programme
(Grant Agreement No. 678862), and CRC 183 of the Deutsche
Forschungsgemeinschaft.
M. R. is further grateful to the Brown Investigator Award, a program of
the Brown Science Foundation, the University of Washington College of Arts and Sciences, and the Kenneth K.
Young Memorial Professorship for support. F.N.  acknowledges support from the U.S. Department of Energy, Office of Science, Basic Energy Sciences under award  DE-SC0019166, the Simons Foundation under award 623768.
The work presented here is supported by the Carlsberg Foundation, grant CF22-0727.


\appendix 

\section{Expressions for  $\cV  $} 
\label{app:lindblad_frame_transf_expr}
Here we provide an explicit expression for the superoperator $
\delta \cV $ which enters in the  memory dressing transformation.
We first show how $\delta \cV$ in Eq.~\eqref{eq:M_timedomain} is obtained from the corresponding expression given in Ref.~\cite{Nathan_2020b}. 
Subsequently (for cases with time-independent Hamiltonians) we provide complementary expressions for $\delta \mathcal V$ in terms of the eigenbasis of the Hamiltonian (Sec.~\ref{seca:eigenbasis}) and in terms of an iterative expansion using nested commutators with the Hamiltonian (Sec.~\ref{seca:taylor_series}).
The latter expression may be used in cases where diagonalization of the Hamiltonian is not feasible. 
\begin{widetext}


As defined in Ref.~\cite{Nathan_2020b},  the interaction picture version of $\delta \cV$, $\delta \tilde{\cV}(t)$, is given  by~\cite{this_expr}:
\newcommand{\dabc}{{ d}a\,{ d}b\,{ d}c\,}
\newcommand{\dab}{{d}a\, { d}b\,}
\newcommand{\da}{{ d}a\,}
\newcommand{\dom}{{ d}^2\omega\,}
\be 
\delta \tilde{\cV}(t)[\rho] = \int_{-\infty}^\infty \!\! \da\int_{-\infty}^\infty \!\! db\, [\theta(a-t)-\theta(b-t)]\mathcal G(a,b)[\rho],
\label{eqa:mexpr}\ee
where $\mathcal G(a,b)[\rho]$ is a superoperator whose action on a generic density operator $\rho$ is given by
\be 
\mathcal G(a,b)[\rho] = -\gamma \int_{-\infty}^a\!\!\!\!\! {d}c\, g^*(a-b)g^*(b-c)[\tilde X(a), \rho\tilde X(c) ]+{\rm h.c.}.
\ee 
Thus we may write 
\be 
\delta \tilde{\cV}(t)[ \rho] = \gamma \int_{-\infty}^\infty\!\!\! da\,\int_{-\infty}^\infty\!\! dc\, [\tilde X(a),\rho \tilde X(c)] f(a-t,c-t)  + {\rm h.c.},
\label{eqa:m_def_1}
\ee
with $f(t,t')$ defined below Eq.~\eqref{eq:M_timedomain} in the main text: 
\be 
f(t,t') \equiv \int_{-\infty}^\infty \!\! db\, 
g^*(t-b)g^*(b-t')\theta(b-t')[\theta(t)-\theta(b)].
\ee 
When $H_{\mathcal S}$ is time-independent, the Schr\"odinger picture version of $\delta \tilde{\cV}$ can be obtained by setting $t=0$: $\delta {\cV}= \delta \tilde{\cV}(0)$. 
Doing this above yields the result quoted in Eq.~\eqref{eq:M_timedomain}. 

\subsection{Expression for $\mathcal V$ in energy eigenbasis}
\label{seca:eigenbasis}
When $H_{\mathcal S}$ is time-independent we  can conveniently  express ${\mathcal V}$ in terms of the energy eigenbasis: 
writing $\tilde X(t) =  \sum_{mn}   X_{mn} e^{-i \omega_{mn} t}$, where $\omega_{mn}=E_m-E_n$, $ X_{mn} \equiv |m\rangle\langle m| X|n\rangle \langle n|$,    and $H_{\mathcal{S}}\Ket{n} = E_n\Ket{n}$, gives us  
\be 
\delta {\cV}[ \rho] =   \sum_{m,n,k,l} [ X_{mn}, {\rho}  X_{kl}^\dagger]c_{mn;kl}+ {\rm h.c.},\!\!
\ee 
with $c_{mn;kl} = 4\pi^ 2f(-\omega_{mn},\omega_{kl})$.
Here $f(\omega,\omega') =\frac{1}{4\pi^2}\int_{-\infty}^\infty  dt \int_{-\infty}^{\infty} dt' e^{i\omega t}e^{i\omega' t'}f(t,t')$ denotes the Fourier transform of $f(t,t')$.
A straightforward evaluation of the Fourier transform yields
\be 
c_{mn;kl} = 2\pi\gamma\!\int_{-\infty}^\infty \!\! dq \,  \frac{g(q)[g(q)-g(q+\omega_{mn}-\omega_{kl})]}{(\omega_{mn}-\omega_{kl})(q-\omega_{mn}-i0^+)}, 
\ee 
where, for $\omega_{mn}=\omega_{kl}$, 
the integrand  
should be evaluated using L'Hospital's rule.

\subsection{Expression of $\cV$ in terms of commutator expansion}
\label{seca:taylor_series}
In cases where diagonalization of the Hamiltonian is not feasible, $\cV$ can be computed using an iterative series expansion akin to the one provided for the jump operator in Ref.~\cite{Nathan_2020b}. 
We review this expansion for the case of a single bath and a time-independent Hamiltonian, while noting that our results can be extended to multiple baths and time-dependent Hamiltonians. 
Our first step is to write {an expansion for the interaction picture operator $\tilde{X}(t)$}:
\be 
\tilde X(t)  = \sum_{n=0}^\infty \frac{X^{(n)} t ^n}{n!}+ {\rm h.c.}, \quad {X^{(n)} = -i[H_{\mathcal S},X^{(n-1)}]}.
\ee 
Here $X^{(n)}$ is the operator obtained from $X$ after $n$ commutation operations with $-iH_{\mathcal S}$, {where we define 
$X^{(0)} \equiv X$}. 
Using this {expression} in the definition of $\cV$ [Eq.~\eqref{eq:M_timedomain} of the main text] yields 
\be 
{\cV}[\rho] = \gamma \sum_{m,n=0}^\infty 
[X^{(m)},\rho X^{(n)}] K_{mn} +{\rm h.c.}, \quad
\label{eqa:taylor_series}
K_{mn} \equiv \frac{1}{m! n!} \int_{-\infty}^\infty\!\!\! dt\int_{-\infty}^\infty\!\! dt'\, f(t,t') t^m t'{}^n.
\ee

Referring back to Eq.~\eqref{eqa:mexpr} with $t$ set to zero {(to obtain $\mathcal{V}$ in the Schr\"{o}dinger picture)}, and noting that $\mathcal G(a,b)$ decays to zero with $|a|,|b|$, we see that $\mathcal V$ itself can be  computed with a temporal cutoff with an error that goes to zero as the cutoff is increased.
For any finite value of the cutoff, the corresponding integrals in Eq.~\eqref{eqa:taylor_series} are finite, and the sum converges.
The coefficients $\{K_{mn}\}$ are  inexpensive to compute and depend only on the bath jump correlator (but not on any details of the system).
{Thus the expression in Eq.~(\ref{eqa:mexpr}) provides a viable way to calculate $\mathcal{V}$ to good accuracy for systems where exact diagonalization is not feasible.}

\end{widetext} 
\twocolumngrid
\section{Relationship between ${\bar \rho_{\rm ULE}}$ and $\bar \rho'_{\rm exact}$}
\label{app:exact_and_ule}
Here we establish that $\bar \rho'_{\rm exact} = {{\bar{\rho}_{\rm ULE}}}  -\mathcal L_{\rm ULE}^{-1}[\bar  \xi]$. The result is quoted above Eq.~(13) in the main text.
As our starting point, we 
write a formal solution of Eq.~(\ref{eq:ule_exact}) of the main text as 
\be 
\rho'(t) = e^{\mathcal L _{\rm ULE} t} [\rho'(0)]+ \int_{0}^t ds\,  e^{\mathcal L_{\rm ULE}(t-s)}[ \xi(s)].
\label{eqa:ule_solution}´
\ee
We obtain the exact steady state of the system (in the memory-dressed frame) by taking the $t\to \infty$ limit above: $\lim_{t \to \infty}\rho'(t) = \bar \rho'_{\rm exact}$. 

To compute the $t\to \infty$ limit on the right-hand side, we consider the eigendecomposition of $\mathcal L_{\rm ULE}$. 
Due to its  Lindblad form and our assumption of a unique steady state, the superoperator  $\mathcal L_{\rm ULE}$ has a single vanishing eigenvalue, while all other eigenvalues have strictly negative real parts. 
The left and right eigenvectors corresponding to the vanishing eigenvalue  are the identity operator and {${\bar \rho_{\rm ULE}}$}, respectively. 
{(As in the main text, we have defined {$\mathcal{L}_{\rm ULE}[\bar{\rho}_{\rm ULE}] = 0$}.)}
Hence $\lim_{t \to \infty} e^{\mathcal L_{\rm ULE} t}[{ \mathcal O}] = \Tr[{\mathcal O}]\, {\bar{\rho}_{\rm ULE}} $.
Using $\Tr[\rho']=1$ we thus conclude 
$ 
\lim_{t \to \infty} e^{\mathcal L _{\rm ULE} t} [\rho'(0)] = {{\bar \rho_{\rm ULE}}}.
$ 
On the other hand, Eq.~\eqref{eq:ule_exact} of the main text implies that $\Tr[ \xi(t)]=0$; hence   $e^{\mathcal L _{\rm ULE} ( t-s)} [\xi(s)]$ must decay exponentially with $t-s$. 
Taking the limit $t\to \infty$ in  Eq.~\eqref{eqa:ule_solution} hence gives 
\be 
\bar \rho'_{\rm exact} = {{\bar{\rho}_{\rm ULE}}}  -\mathcal L_{\rm ULE}^{-1}[\bar  \xi],
\ee
where $ \mathcal L_{\rm ULE}^{-1}[\bar \xi] \equiv {-}\int_{0}^\infty  ds\, e^{\mathcal L_{\rm ULE}s}\bar  \xi $ is finite due to the fact that $\bar \xi$ is traceless. 
This is the result we wished to establish.

\section{Scaling of $C_\cO$}
\label{app:co_scaling}
Here we show that the dimensionless constant $C_\cO$ [first appearing in Eq.~(\ref{eq:ltf_expval}) of the main text] is generically  finite in the limit $\Gamma \to 0$. 
To this end, we recall the definition of {$C_{\mathcal{O}}$}: 
\be 
C_\cO = \frac{2 \Gamma  \norm{\mathcal L^{\dagger -1}_{\rm ULE}[\cO'-\langle \cO'\rangle_{\rm ULE}]}}{|\langle \bar \cO'\rangle_{\rm ULE}|}.
\label{eqa:ca_def}
\ee 
In the following, we use $\Theta(\Gamma^n)$ to indicate quantities that scale as $\Gamma^n$ in the limit of small $\Gamma$. Likewise, we use $O(\Gamma^n)$ to indicate quantities that scale as $\Gamma^n$ or slower in the same limit. 
Our goal is to show that $C_\cO= O(\Gamma^0)$ for generic operators $\cO$.


We first consider  the case $\cO = S$ where  ${\langle S\rangle_{\rm ULE}}$ is nonzero in the limit $\Gamma \to 0$. 
Since we assume the system to have a unique steady state,   $\mathcal L_{\rm ULE}^\dagger$ has a single vanishing eigenvalue with corresponding left eigenvector {$\bar \rho_{\rm ULE}$}; all other eigenvalues either scale as $\Theta(\Gamma)$ or $\Theta(\Gamma^0)$. 
Thus, for any 
($\Gamma$-independent) operator ${\mathcal M}$ that is orthogonal to {$\bar{\rho}_{\rm ULE}$} (in the sense of the Hilbert-Schmidt inner product), 
we infer that $\mathcal L^{\dagger -1}_{\rm ULE}[{\mathcal M}] = O(\Gamma^{-1})$. 
Since $S'-\langle S'\rangle_{\rm ULE}$ by definition has zero expectation value in the ULE steady state and is hence orthogonal to {$\bar{\rho}_{\rm ULE}$},  we  have $\mathcal L^{\dagger -1}_{\rm ULE}[S'-\langle S'\rangle_{\rm ULE}] =O(\Gamma^{-1})$. 
Using this in Eq.~\eqref{eqa:ca_def}, we conclude that $C_S=O(\Gamma^0)$ for observables with finite expectation values in the limit $\Gamma \to 0$, i.e., with $\lim_{\Gamma \to 0} {\langle S'\rangle_{\rm ULE}} \neq 0$. 

Next, we consider the case $\cO = I$, where $I$ has vanishing steady-state value in the limit $\Gamma \to 0$. 
Since  $\lim_{\Gamma \to 0}\bar \rho'$ is diagonal in the eigenbasis  of the Hamiltonian~\cite{Nathan_2020b}, there are two classes of operators for which $\lim_{\Gamma \to 0}{\langle \cO'\rangle_{\rm ULE}} = 0$: 
the first class are operators whose diagonal matrix elements vanish in the eigenbasis of $H_{\mathcal S}$. 
The second class of operators are those that have nonzero diagonal elements in the eigenbasis of $H_{\mathcal S}$, but whose combination of diagonal elements nevertheless causes $\cO$ to be  orthogonal to $\bar \rho_{\rm ULE}$.
For the  latter  class of operators, $C_\cO$ can be  infinite. 
However, the vanishing steady-state value of these   operators requires fine-tuning of the system and bath parameters -- the operator will  generically acquire nonzero steady-state values under perturbations of the bath parameters (such as temperature or chemical potential) or system-bath coupling. 
We hence expect this case to be  non-generic. 

Operators whose diagonal matrix elements vanish in the eigenbasis of the Hamiltonian  include many physically relevant quantities,  such as the  total currents of conserved quantities and,  in cases of Hamiltonians with time-reversal symmetry, current densities. 
For  this class of operators, we can write  
\be 
I = -i[H_{\mathcal S},Q],
\label{eqa:qa_result}
\ee 
for some finite operator $Q$~\cite{note_that_Q}.
The magnon current {$I_{n+1,n}$} for the spin chain we consider in the main text can for example be written in the form above with $Q =\sum_{k=1}^n \sigma_k^z$ (i.e., with $Q$ measuring the 
{total $z$-component of spin}
on sites $1\ldots n$).
We now use the above property to infer the scaling behavior of $C_I$. 

First, we {rewrite Eq.~(\ref{eqa:qa_result}) as} $I = {\mathcal{L}^\dagger_{\rm ULE}}[Q]-\mathcal D_{\rm ULE}^\dagger[Q]$, where $\mathcal D_{\rm ULE}$ denotes the bath-induced component of the ULE Liouvillian {(including the Lamb shift)}. 
Writing $\cV^{\dagger -1}[{{I}}] =  I  + \left(  \frac{1}{1+\delta \cV^\dagger}-1\right)[I]$, and using $I' =\cV^{\dagger -1}[{{I}}]$, we thus have 
\be 
I'
= \mathcal L^\dagger_{\rm ULE}[{Q}] +{R},
\label{eqa:ApR}\ee 
with 
\be 
R = - \mathcal D^\dagger_{\rm ULE}  [Q] + \left(\frac{1}{1+\delta \cV^\dagger}-1\right)[I].
\ee
We now note that  $\mathcal L_{\rm ULE}[{\bar{\rho}_{\rm ULE}}] = 0$, implying  $\Tr({\bar{\rho}_{\rm ULE}} \mathcal L_{\rm ULE} ^\dagger [Q])= 0$.
{Using this relation with Eq.~(\ref{eqa:ApR}) and the definition $\Avg{I'}_{\rm ULE} = {\rm Tr}[\bar{\rho}_{\rm ULE}\frac{1}{1+\mathcal \delta \cV^\dagger}{I}]$, we find}
\be 
{\langle I'\rangle_{\rm ULE} = \Tr[\bar{\rho}_{\rm ULE} R]}.
\label{eqa:AbarR}
\ee

To infer the scaling behavior of $R$, we first use  $\mathcal D_{\rm ULE}^\dagger [Q] =  \Theta( \Gamma)$~\cite{Nathan_2020b}. 
Moreover, since $\norm{\delta \mathcal V}\leq \Gamma \tau$, we have $\left((1+\delta \cV^\dagger)^{-1}-1\right)[I]  =  \Theta(\Gamma)$. 
Thus 
$ 
R =  \Theta(\Gamma)  
$,
implying
\be 
\langle I'\rangle_{\rm ULE} = \Theta(\Gamma).
\ee 
This demonstrates that $\lim_{\Gamma \to 0}\langle \bar A\rangle = 0$ for operators with vanishing diagonal matrix elements in the eigenbasis of $H_{\mathcal S}$, as we inferred above.

We now show that $\mathcal L^{\dagger -1}[I'-\langle I'\rangle_{\rm ULE}] =\Theta(\Gamma^0)$ for operators that can be written in the form in Eq.~\eqref{eqa:qa_result}. 
Using Eq.~(\ref{eqa:ApR}), we first note that 
\be
\mathcal L_{\rm ULE}^{\dagger -1}[I'-\langle I'\rangle_{\rm ULE}] =     Q + \mathcal L^{\dagger -1}_{\rm ULE}[R-\langle I'\rangle_{\rm ULE}].\vspace{0.075 in}
\ee

To analyze the second term on the right hand side of this expression, we note two important facts.
First, since  $\langle I' \rangle_{\rm ULE}= \Tr[\bar{\rho}_{\rm ULE} {R}]$ [see Eq.~(\ref{eqa:AbarR})],   $R-{\langle I'\rangle_{\rm ULE}}$ is  orthogonal to ${\bar{\rho}_{\rm ULE}}$. 
Moreover, according to the discussion above, $R-{\langle I'\rangle_{\rm ULE}} =  \Theta(\Gamma)$. 
Therefore $\mathcal L^{\dagger -1}_{\rm ULE}[{R}-{\langle I'\rangle_{\rm ULE}}] =  \Theta (\Gamma^{0})$.
Using $Q = \Theta(\Gamma^0)$, we find 
\be 
\norm{\mathcal L^{\dagger -1}[{I'-\langle I'\rangle_{\rm ULE}}]}  =\Theta  (\Gamma^0). 
\ee 
Combining this with ${\langle \bar I'\rangle_{\rm ULE}} = \Theta (\Gamma)$ in Eq.~\eqref{eqa:ca_def}, we conclude $C_I$ is finite in the limit $\Gamma \to 0$.

\section{Steady-state value of magnetization}
\label{app:magnetization}

\begin{figure}[b]
   \centering
   \includegraphics[width=0.99\columnwidth]{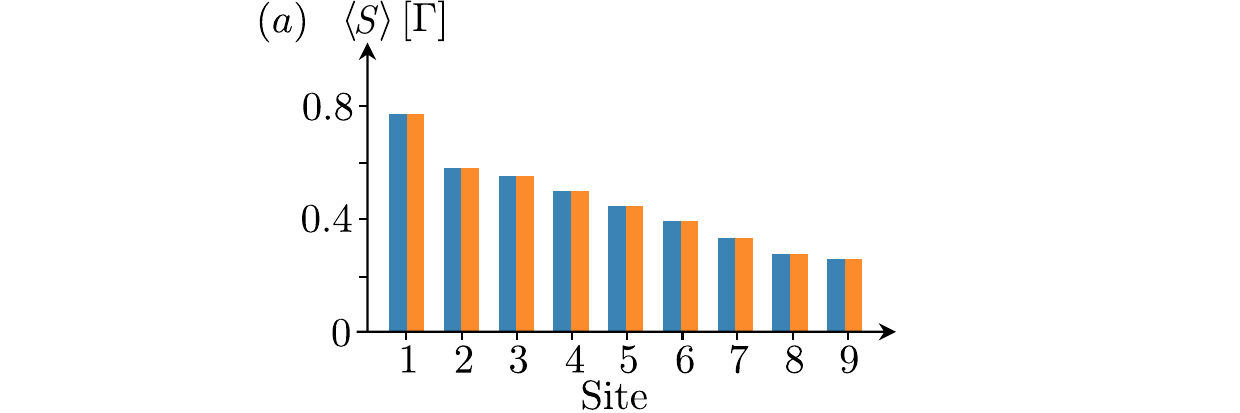}
   \caption{Steady-state values of the undressed ($\{\Avg{{\sigma}_n^z}_{\rm ULE}\}$, blue) and dressed ($\{\Avg{{\sigma}_n^z{}'}_{\rm ULE}\}$, orange) on-site magnetizations in the spin-chain model considered in the main text.}
   \label{fig:fig_3}
\end{figure}

Here we provide data for the steady-state values of magnetization on each site, $\{\sigma_n^
z\}$, for the $9$-site spin chain model considered in the main text.
In Fig.~\ref{fig:fig_3} we show the steady-state values of the magnetization obtained with (orange) and without (blue) applying the correction from the  memory dressing transformation, $\langle  \sigma_n^z{}'\rangle_{\rm ULE}$, and $\langle  \sigma_n^z\rangle_{\rm ULE}$, respectively.
As for the computation of the bond current in the main text, we apply the inverse memory-dressing transformation through expansion to first order, setting $\cV^{-1}\approx (1-\delta \mathcal V)$ to compute $\langle  \sigma_n^z{}'\rangle_{\rm ULE}$. 
The two data sets coincide up to negligible relative corrections, {as we expected due to the finite value of $\Avg{{\sigma}_n^z}_{\rm ULE}$ in the weak coupling limit (see main text).}
\bibliography{bibliography_master_equation}

\begin{thebibliography}{50}
\bibitem [{sbn()}]{sbnorm}%
  \BibitemOpen
  \href@noop {} {}\bibinfo {note} {A nonzero expectation value can be absorbed into $H_{\mathcal S}$ after appropriate redefinition of $B$. If $X$ has an unbounded spectrum, an appropriate cutoff for its eigenvalues has to be implemented.}\BibitemShut {Stop}%
\bibitem [{uni()}]{unique_fixed_point}%
  \BibitemOpen
  \href@noop {} {}\bibinfo {note} {We consider systems which have a unique steady state; cases with multiple fixed points, such as systems that support spontaneous symmetry breaking or local integrals of motion, are left for future studies}\BibitemShut {NoStop}%
\bibitem [{tra()}]{trace_norm_bound}%
  \BibitemOpen
  \href@noop {} {}\bibinfo {note} {In Ref.~\cite{Nathan_2020b}, the bound was established for the spectral norm; however the same arguments are valid for the trace norm.}\BibitemShut {Stop}%
\bibitem [{inf()}]{infca}%
  \BibitemOpen
  \href@noop {} {}\bibinfo {note} {One can construct operators with nonvanishing diagonal matrix elements in the energy basis, whose expectation values vanish in the limit $\Gamma\to 0$ (such that the relative deviation does not scale to zero with $\Gamma$). We expect that such observables are non-generic and require fine-tuning.}\BibitemShut {Stop}%
\bibitem [{fn_()}]{fn_bond_current_uniformity}%
  \BibitemOpen
  \href@noop {} {}\bibinfo {note} {We emphasize that uniformity of the bond currents alone does not imply that the value of current is accurate. For example, while the BR approach ensures uniformity of the current by construction, its accuracy is guaranteed only in the weak coupling limit~\cite{BRBound}.}\BibitemShut {Stop}%
\bibitem [{inv()}]{invlft}%
  \BibitemOpen
  \href@noop {} {}\bibinfo {note} {We apply the inverse frame transformation through expansion to first order, setting ${\mathcal{V}}^{-1}\approx 1-\delta\mathcal V$}\BibitemShut {NoStop}%
\bibitem [{not()}]{note_that_Q}%
  \BibitemOpen
  \href@noop {} {}\bibinfo {note} {Note that $Q$ is not uniquely defined, since adding a term diagonal in the eigenbasis of $H_{\mathcal S}$ to $Q$ will not change $I$.}\BibitemShut {Stop}%
\bibitem [{BRB()}]{BRBound}%
  \BibitemOpen
  \href@noop {} {}\bibinfo {note} {Specifically, $ \left|\langle \cO\rangle_{\rm exact} -\langle \cO\rangle_{\rm BR}\right| \leq \Gamma\tau C_{\cO} |{\langle \bar {\cO}\rangle_{\rm BR}}|, $ with $C_{\cO}' \equiv \Gamma \norm{(\mathcal L_{\rm BR}^{\dagger})^{-1}[{\cO}-\langle \cO\rangle_{\rm BR}]}{/\Avg{\cO}_{\rm BR}}$, where $\langle \cO\rangle_{\rm BR}$ denotes the expectation value of $\cO$ in the steady state of the BR equation and $\mathcal L_{\rm BR}$ denotes the Bloch-Redfield Liouvillian. For small $\Gamma$, $C'_\cO$ follows similar scaling behavior to that of $C_\cO$ as described in the SOM~\cite{SOM}.}\BibitemShut {Stop}%
\end{thebibliography}

\end{document}